\documentclass[12pt]{article}

\usepackage[super,sort&compress,comma]{natbib} 

\usepackage[left=2.5cm, right=2.5cm, top=3cm, bottom=3cm]{geometry}

\usepackage{graphicx} 

\usepackage[english]{babel}

\usepackage[T1]{fontenc}
\usepackage{hyperref}
\usepackage[utf8]{inputenc}
\usepackage{amsmath}

\title{Tailoring Intercalant Assemblies at the \\ Graphene--Metal Interface}
\author{
\noindent
Johannes Halle$^{\ast}$, Nicolas Néel, Jörg Kröger \\
\small{Institut für Physik, Technische Universität Ilmenau, D-98693 Ilmenau, Germany} \\ 
\small{E-mail: johannes.halle@tu-ilmenau.de} \\
\scriptsize \textit {This document is the unedited Author's version of a Submitted Work that was subsequently accepted for} \\ 
\scriptsize \textit {publication in Langmuir, copyright \copyright American Chemical Society and the Division of Chemical Education, Inc.,} \\
\scriptsize \textit {after peer review. To access the final edited and published work see} \\
\scriptsize \textit {\href{https://pubs.acs.org/articlesonrequest/AOR-k9bCzvtT3V6WRd3q4UUw}{https://pubs.acs.org/articlesonrequest/AOR-k9bCzvtT3V6WRd3q4UUw}} \\
}
\date{}

\begin{document}

\maketitle

\begin{abstract}

The influence of graphene on the assembly of intercalated material is studied using low-temperature scanning tunneling microscopy.
Intercalation of Pt under monolayer graphene on Pt(111) induces a substrate reconstruction that is qualitatively different from the lattice rearrangement induced by metal deposition on Pt(111) and, specifically, the homoepitaxy of Pt. 
Alkali metals Cs and Li are used as intercalants for monolayer and bilayer graphene on Ru(0001)\@.
Atomically resolved topographic data reveal that at elevated alkali metal coverage $(2\times 2)$Cs and $(1\times 1)$Li intercalant structures form with respect to the graphene lattice.

\end{abstract}

\section*{Introduction}

Superstructures at surfaces are ubiquitous. \cite{springer_2006,springer_2015}
They may occur as reconstructions of the clean surface, as an ordered assembly of adsorbates, as a moiré pattern due to the lattice mismatch between adsorbate and substrate lattice or as an electronic superlattice.

In the case of adsorbates, superlattices form as the result of the balance between adsorbate--adsorbate and adsorbate--substrate interactions.
Exemplarily, when the adsorbate--adsorbate interaction is dominant characteristic molecular patterns may occur, \cite{nature_437_671,arpc_58_375,aipadv_2_041402} while a stronger adsorbate--substrate coupling may be exploited for template effects, such as the guided adsorption on vicinal surfaces, \cite{prb_56_2340,prb_61_2254,apl_78_829,epl_58_730,jpcm_15_s3311,am_18_174,apl_88_163101,ss_601_4180,jpcs_187_012025} on molecular platforms, \cite{acie_39_1230,nature_424_1029} and on moiré lattices. \cite{science_303_217,science_319_1824,acie_46_5115,langmuir_23_2928,jpcc_112_8147,jpcc_118_13320,jacs_131_14136,njp_11_103045,apl_95_093106,jcp_131_164701,acsnano_6_3034,apl_96_093115,acie_49_1794,ss_605_1676,jacs_133_9208,acsnano_6_944,acsnano_7_11341,acsnano_7_11121,nl_13_3199,jpcc_120_8772,nl_16_7610,jpcc_121_1639,prb_98_115417} 
Adsorbate--substrate interactions mediated by surface state electrons \cite{science_274_118,apa_66_1107,prl_85_2981,prb_65_115420,jpcm_12_l13,prl_92_016101} and electronic moiré patterns \cite{cpl_484_59} were demonstrated to steer the adsorbate assembly as well.

The impetus to the work presented here was the addition of a third interaction and the exploration of a possible further tailoring of surface structures.
To this end, graphene-covered metal surfaces, Pt(111) and Ru(0001), and the intercalation of Pt and alkali metals (Cs, Li) were used.
The intercalated material experiences the coupling to the substrate as well as to the graphene.
Pt intercalation under graphene on Pt(111) leads to a reconstruction of the Pt surface, which is related to the structures reported for homoepitaxial growth of Pt on pristine Pt(111)\@.
However, the presence of graphene induces a considerably extended long-range order of the reconstruction as well as a qualitative change in the underlying dislocation network.
For Cs and Li intercalation on graphene-covered Ru(0001) alkali metal assemblies form with respect to the graphene lattice rather than to the metal substrate.

\section*{Experimental Methods}

The experiments were performed in ultrahigh vacuum ($10^{-9}\,\mathrm{Pa}$) with a scanning tunneling microscope (STM) operated at $5\,\text{K}$\@.
Surfaces of Pt(111) and Ru(0001) were cleaned by repeated Ar$^+$ bombardment and annealing at $1200\,\text{K}$ in O$_2$ atmosphere ($4\cdot 10^{-5}\,\text{Pa}$)\@. 
Graphene was prepared on the clean surfaces by thermal decomposition of C$_2$H$_4$\@. 
On Ru(0001) a second layer of graphene was formed by segregation of bulk C\@. \cite{natmater_7_406} 
Metal intercalation was performed by, first, depositing Pt from a hot filament, Cs and Li from commercial dispensers onto the respective metal surfaces at room temperature and, second, by annealing the sample at $1200\,\text{K}$ (Pt), $660\,\text{K}$ (Cs), $570\,\text{K}$ (Li)\@.
Using this preparation protocol leads to the efficient intercalation of the deposited material, as previously shown for Cs, \cite{natcommun_4_2772} Li, \cite{jpcc_120_5067,nl_18_5697} and Ni. \cite{nl_18_5697}
The annealing after deposition is particularly important for Cs and Pt since otherwise adsorbed clusters remain on graphene. \cite{apl_95_093106,natcommun_4_2772,prb_93_045426}  
The intercalation is further evidenced by the concomitant weakening of the graphene moiré pattern as well as by STM images showing the atomically resolved graphene lattice.
Cointercalation proceeded \textit{via} the deposition of Li onto Cs-intercalated graphene on Ru(0001), followed by annealing.
STM tips were fabricated from pure Au wire and trained \textit{in situ} by annealing and field emission on a Au substrate. 
Topographic data were acquired with constant current and the bias voltage applied to the sample. 
STM images were processed using WSxM. \cite{rsi_78_013705}

\section*{Results and Discussion}

\subsection*{Pt(111)}

Annealing Pt(111) at temperatures exceeding $1330\,\text{K}$ induces a hexagonal reconstruction of the surface. \cite{prl_68_2192,prb_48_18119} 
STM studies showed that exposing Pt(111) to Pt vapor stabilizes a similar reconstruction already at $400\,\text{K}$, which resembles a honeycomb network of protruding lines. \cite{prl_70_1489}
These characteristic superstructures consist of domains where the bulk face-centered cubic (fcc) stacking is retained alternating with hexagonal close-packed (hcp) stacking regions.
More recent calculations using the Frenkel-Kontorova model \cite{pzs_13_1} showed that Pt(111) indeed teeters at the brink of a stability domain and may be triggered to reconstruct by small environmental changes. \cite{prb_67_205418}
The Pt(111) reconstruction is related to the soliton reconstruction of Au(111) \cite{prl_54_2619,prb_39_7988,prb_42_9307,prl_69_1564} and likewise occurs upon deposition of Co, \cite{ss_337_147} Cr, \cite{prb_57_R4285} Cu \cite{prb_58_R10195} on Pt(111)\@.

\begin{figure}
\begin{center}
\includegraphics[width=0.75\linewidth]{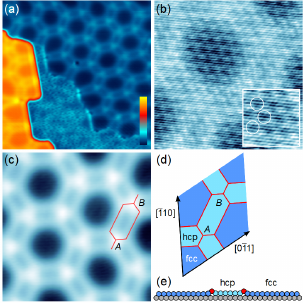}
\end{center}
\caption{(Color online) 
Reconstruction of graphene-covered Pt(111) after Pt intercalation.
(a) STM image of Pt-intercalated graphene on Pt(111) (bias voltage $V=1\,\text{V}$, tunneling current $I=100\,\text{pA}$, size: $40\times 40\,\text{nm}^2$)\@.
Two adjacent Pt(111) terraces are visible.
The honeycomb network is due to a reconstruction of the Pt(111) surface.
(b) Atomically resolved close-up view of (a) showing the graphene lattice with parts of the reconstruction ($50\,\text{mV}$, $100\,\text{pA}$, $12\times 12\,\text{nm}^2$)\@.
Inset: Close-up view of (b) with circles indicating a moiré pattern ($3.5\times 3.5\,\text{nm}^2$)\@.
(c) STM image of the Pt(111) reconstruction network showing the characteristic fcc (hexagonal dark depressions), hcp (stripelike depressions enclosed by bright double lines) stacking domains ($1\,\text{V}$, $100\,\text{pA}$, $19\times 19\,\text{nm}^2$)\@.
The triangle-shaped bright protrusions are the connectors of the network (see text)\@.
(d) Illustration of structural elements of the reconstruction network.
Regions with hcp stacking are separated from fcc domains by pairs of lines that are oriented along $\left[ 11\bar{2}\right]$, $\left[ 1\bar{2}1\right]$, $\left[\bar{2}11\right]$ directions.
The double lines intersect at \textit{A} and \textit{B}\@.
(e) Side view of first and second Pt(111) atomic layer along $\langle\bar{1}10\rangle$\@.
$30$ surface Pt atoms reside atop $29$ second-layer Pt atoms giving rise to fcc and hcp stacking regions with protruding double lines.
}
\label{fig1}
\end{figure}

Here, possible modifications of this reconstruction due to the presence of graphene are investigated.
Figure \ref{fig1}a shows a representative STM image of two adjacent terraces of graphene-covered Pt(111) after Pt intercalation.
Most obvious is a regular honeycomb superstructure that extends over a few hundred nanometers.
It is characterized by a spatial period of $7.99\pm 0.08\,\text{nm}$ with parallel ridges separated by $1.98\pm 0.10\,\text{nm}$.
This superstructure is assigned to a reconstruction of the Pt(111) surface due to its resemblance to previously reported reconstructions on that surface. \cite{prl_68_2192,prl_70_1489}
In particular, the ridges will be referred to as double lines following conventional phrasing. \cite{prl_70_1489,ss_337_249}.
Deviating from the rather irregular network of double lines observed from the reconstruction induced by the homoepitaxial growth of Pt on clean Pt(111) \cite{prl_70_1489,ss_337_249} the presence of graphene apparently causes a regular and extended honeycomb reconstruction pattern.
Before analyzing the superstructure in more detail two remarks are noteworthy.
First, the STM image of Figure \ref{fig1}a shows that regions without reconstruction may occur.
A surface area on the lower terrace and close to a substrate edge does not exhibit a reconstructed surface.
Second, the presence of monolayer graphene (MLG) can be inferred from the atomically resolved graphene lattice (Figure \ref{fig1}b)\@.
As indicated by the circles in the inset to Figure \ref{fig1}b, a moiré pattern with a spatial period of $1.00\pm 0.02\,\text{nm}$ is visible.
The angle between the moiré and graphene lattice is $1.2^\circ\pm 1.1^\circ$, which is in accordance with the angle of $14.5^\circ\pm 0.7^\circ$ enclosed by graphene and the Pt(111) lattice.
This graphene lattice orientation is observed on both reconstructed and unreconstructed surface regions since graphene spans these regions without discontinuity (Figure \ref{fig1}a)\@.  
A second orientation was observed on Pt(111) with a smaller spatial period of $0.79\pm 0.02\,\text{nm}$ and an angle of $-17.5^\circ\pm 0.8^\circ$ enclosed by the graphene and Pt lattices.
These moiré patterns are characteristic for MLG on Pt(111)\@. \cite{acsnano_5_5627,apl_98_033101} 
The rather low number of graphene orientations observed here is probably due to the rather high annealing temperature used for the intercalation of the Pt atoms, which is in agreement with a previous report. \cite{acsnano_5_5627}

Figure \ref{fig1}c presents topographic data that unravel the structural elements of the observed reconstruction.
The illustrations in Figures \ref{fig1}d,e help identify crystallographic directions and surface regions with different stackings.
The double lines represent a pair of Shockley partial dislocations that separate hexagonal fcc stacking domains from smaller and elongated hcp-stacked regions. \cite{prl_70_1489,ss_337_249}
Thus, across a pair of Shockley partial dislocations one additional Pt atom is incorporated into the surface, which increases the surface atom density.
In the one-dimensional illustration (Figure \ref{fig1}e) $30$ surface Pt atoms reside on top of $29$ second-layer Pt atoms along $\langle \bar{1}10 \rangle$ directions.

The arrangement of double lines in the honeycomb reconstruction leads to an isotropic compression of the Pt(111) surface.
Previously, double lines were demonstrated to meet in two different manners leading to two different types of dislocation line connectors, the so-called bright and dark stars. \cite{prl_70_1489,ss_337_249}
Bright stars are protruding intersections of three pairs of double lines, where central Pt atoms reside at on-top sites of Pt(111)\@. 
They provide an energy gain by the annihilation of three point dislocations, which outweighs the energy costs for the accompanying stacking faults. \cite{prl_70_1489,ss_337_249}
Dark stars represent connectors that enclose extended hcp-stacked regions. \cite{prl_70_1489,ss_337_249}
Bright and dark stars reflect the threefold symmetry of Pt(111), which is due to the inequivalence of hcp and fcc sites.
Indeed, in all previous reports the energy difference between hcp and fcc sites established a topological law that forces the two types of connectors -- bright and dark stars -- to alternate at the corners of the honeycomb hexagons. \cite{ss_337_249,prb_67_205418,pzs_13_1}
The experimental findings reported here (Figure \ref{fig1}) invalidate the topological law since the honeycomb reconstruction surprisingly exhibits protrusions at all connector sites of the dislocation network, labeled \textit{A} and \textit{B} in Figure \ref{fig1}c,d.
A possible origin of this observation is discussed next.

The additional interaction of Pt surface atoms with graphene counteracts their coupling to the substrate, which facilitates the reconstruction.  
A similar argument was put forward to explain the stability of the high-temperature Pt/Pt(111) reconstruction.  
In this case, a weaker bonding of the Pt surface layer was attributed to thermally excited vibrations. \cite{prl_68_2192, prb_48_18119, prb_67_205418}
A second important consequence of the attraction between Pt and graphene is the reduction of stacking-dependent energy differences, especially between hcp and fcc sites.  
This aspect is corroborated by the remarkably large ratio of hcp-stacked to fcc-stacked areas, which is $\approx 0.78$ in the case presented here.  
In comparison, the purely temperature-induced reconstruction of Pt(111) at $1330\,\text{K}$ yielded a hcp to fcc ratio of only $0.43$\@. \cite{prl_68_2192,prb_48_18119}
Additional deposition of Pt, however, may further enhance this ratio.  
This was demonstrated in the homoepitaxial growth of Pt on Pt(111) at $400\,\text{K}$ yielding a hcp to fcc ratio of $\approx 0.54$\@. \cite{prl_70_1489,ss_337_249}
Since the energy differences between the different stacking sites are reduced due to the presence of graphene, intralayer interactions become more important. 
Indeed, the strict alternation of bright and dark stars is overcome and the formation of bright stars as connectors of the observed reconstruction network is observed at all six double line intersections of a hexagon.
This hints at a preference of bright stars over dark stars in the presence of graphene due to a lower formation energy.
The above discussion unravels another important aspect that clarifies the role of graphene in the surface reconstruction.
Graphene captures the intercalated Pt close to the substrate.
The presence of Pt vapor was previously demonstrated to facilitate the reconstruction. \cite{prl_70_1489,ss_337_249}

\subsection*{Ru(0001)}

The adsorption of alkali metals on surfaces has a longstanding tradition in surface science. \cite{elsevier_1989,ssr_23_43,jpcm_9_951}
The promotion of catalytic reactions \cite{prb_36_6213,ss_231_280,ss_409_46,ass_140_58} and the increase of electron emission rates \cite{pr_44_423} belong to the appealing alkali-induced effects.
It is therefore not surprising that many aspects of alkali metal adsorption have been studied, \textit{e.\,g.}, the geometric structure of superlattices, \cite{prb_28_6707,prb_45_8638,prb_53_4939,ss_337_198,prb_73_245434,prb_78_245427} vibrational quanta, \cite{prb_53_3658,jpcm_20_224007,ss_449_227,ss_530_170} and lifetimes of electronic excitations \cite{prb_55_10040,prl_82_1931,prl_86_488,apa_78_141,prl_95_176802,pss_82_293,prb_79_075401} 
On graphene, adsorption of alkali metals was shown to, \textit{e.\,g.}, tune the band gap opening between the graphene Dirac cones \cite{acsnano_6_199,prb_92_041407} and modify the electronic transport. \cite{natphys_4_377,apl_111_263502}

More recently, the intercalation of alkali metals on graphene-covered surfaces has moved into a focus of surface science research.
For instance, doping of graphene on Ir(111) may be controlled by the intercalation of Cs. \cite{nl_13_5013}
In addition, the prediction of superconductivity in Li-decorated free-standing graphene \cite{natphys_8_131} has sparked investigations into electronic \cite{prl_104_136803,pnas_112_11795} and structural \cite{jpcc_120_5067} as well as vibrational \cite{nl_18_5697} properties.

\begin{figure}
\begin{center}
\includegraphics[width=0.75\linewidth]{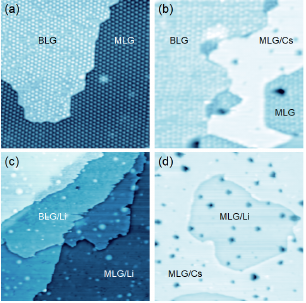}
\end{center}
\caption{(Color online) 
Cs and Li intercalation phases on graphene-covered Ru(0001)\@.
(a) Monolayer (MLG) and bilayer (BLG) graphene on clean Ru(0001) ($200\,\text{mV}$, $100\,\text{pA}$, $120\times 120\,\text{nm}^2$)\@.
The hexagonal superstructure is due to a moiré pattern.
(b) Cs-intercalated graphene (MLG/Cs) between MLG and BLG domains ($200\,\text{mV}$, $80\,\text{pA}$, $90\times 90\,\text{nm}^2$)\@.
(c) Li-intercalated graphene ($300\,\text{mV}$, $100\,\text{pA}$, $300\times 300\,\text{nm}^2$)\@.
Both MLG and BLG are intercalated and labeled as MLG/Li and BLG/Li, respectively.
The intercalated regions in (b) and (c) show an essentially vanishing moiré pattern.
(d) Cointercalation of Cs and Li ($200\,\text{mV}$, $90\,\text{pA}$, $180\times 180\,\text{nm}^2$)\@.
The separate intercalation phases are indicated.
}
\label{fig2}
\end{figure}

In this section, structural aspects of Cs, Li-interclated graphene on Ru(0001) are discussed.
Figure \ref{fig2}a shows an STM image of MLG and bilayer graphene (BLG) on Ru(0001)\@.
The pronounced superlattice is due to a moiré pattern that results from the strong MLG--Ru(0001) hybridization. \cite{pccp_10_3530,prb_78_073401,jcp_130_074705}
The moiré superlattice of MLG exhibits a spatial period of $3.02\pm 0.05\,\text{nm}$, in agreement with previous reports. \cite{prb_76_075429,apl_94_133101,prl_105_236101,apl_104_093110,acsnano_6_9299,njp_18_103027}
The strong MLG buckling of $112\pm 4\,\text{pm}$ mainly reflects the actual topography \cite{pccp_10_3530,nanoscale_4_4687,njp_18_103027} caused by the hybridization of graphene $\pi$-states with Ru $d$-bands. \cite{pccp_10_3530,apl_94_133101}
The angle enclosed by the moiré and MLG lattice is $4.1^\circ\pm 1.0^\circ$, which corresponds to a calculated angle of $0.37^\circ\pm 0.09^\circ$ between the MLG and Ru(0001) lattice.
The MLG spatial period and the moiré twist angles reflect an MLG lattice constant of $0.248\,\text{nm}$, which corresponds to a $0.6\,\%$ tensile stress in MLG\@.
Similar conclusions have been drawn from earlier experimental work and density functional calculations. \cite{pccp_10_3530,prb_76_075429,apl_104_093110,acsnano_6_9299,prl_104_136102,njp_18_103027}

BLG flakes (Figure \ref{fig2}a) exhibit typical diameters of a few $100\,\text{nm}$ and are observed to span several substrate terraces.
Often, they occur at Ru(0001) step edges.
BLG domains display a moiré pattern with a spatial period of $2.95\pm 0.06\,\text{nm}$ and an angle of $1.0^\circ\pm 0.6^\circ$ enclosed with the BLG lattice.
The buckling of BLG is $119\pm 1\,\text{pm}$, which is similar to the corrugation of MLG and in accordance with previous work. \cite{apl_94_133101,nanoscale_4_4687,prl_104_136102,acsnano_6_9299}
In BLG regions, the analysis of the moiré pattern yields a lattice constant of $0.248\,\text{nm}$ and a rotation angle of $0.33^\circ\pm 0.08^\circ$ with respect to Ru(0001) for the lower graphene layer.
Within the uncertainty margins the values are comparable with those obtained for MLG and confirm that the moiré pattern of BLG on Ru(0001) is most likely due to the graphene--Ru(0001) interface. \cite{apl_104_093110,nanoscale_4_4687,pccp_12_5053}
A second moiré pattern was previously reported for BLG on Ru(0001) when elevated C$_2$H$_4$ partial pressures were used for the graphene growth. \cite{acsnano_6_9299,njp_18_103027} 
In the present experiments the C$_2$H$_4$ partial pressure was two orders of magnitude lower than in the previous reports \cite{acsnano_6_9299,njp_18_103027} and, therefore, most likely impeded the formation of the second moiré pattern.

Intercalation of Cs (Figure \ref{fig2}b) and Li (Figure \ref{fig2}c) starts in MLG regions, while BLG intercalation is only observed at saturation coverage (Figure \ref{fig2}c)\@.
In order to determine the corresponding intercalation structures, MLG on Ru(0001) cointercalated by Cs and Li (Figure \ref{fig2}d) is considered.
The cointercalated sample exhibits domains of different apparent heights, where the characteristic buckling of MLG on Ru(0001) is strongly reduced. 
This observation evidences the successful intercalation of graphene with Cs and Li, which efficiently decouple MLG from the Ru surface.
Consequently, the plane domains are attributed to separate Cs and Li intercalation phases, MLG/Cs and MLG/Li.
The phase separation is most likely due to the lower delamination energy of Li-intercalated MLG\@.
It is energetically favorable to form compact Li islands with a decreased graphene--Ru distance compared to the formation of mixed Cs-Li phases where the graphene--Ru separation would be increased due to the larger size of Cs.

\begin{figure}
\begin{center}
\includegraphics[width=0.75\linewidth]{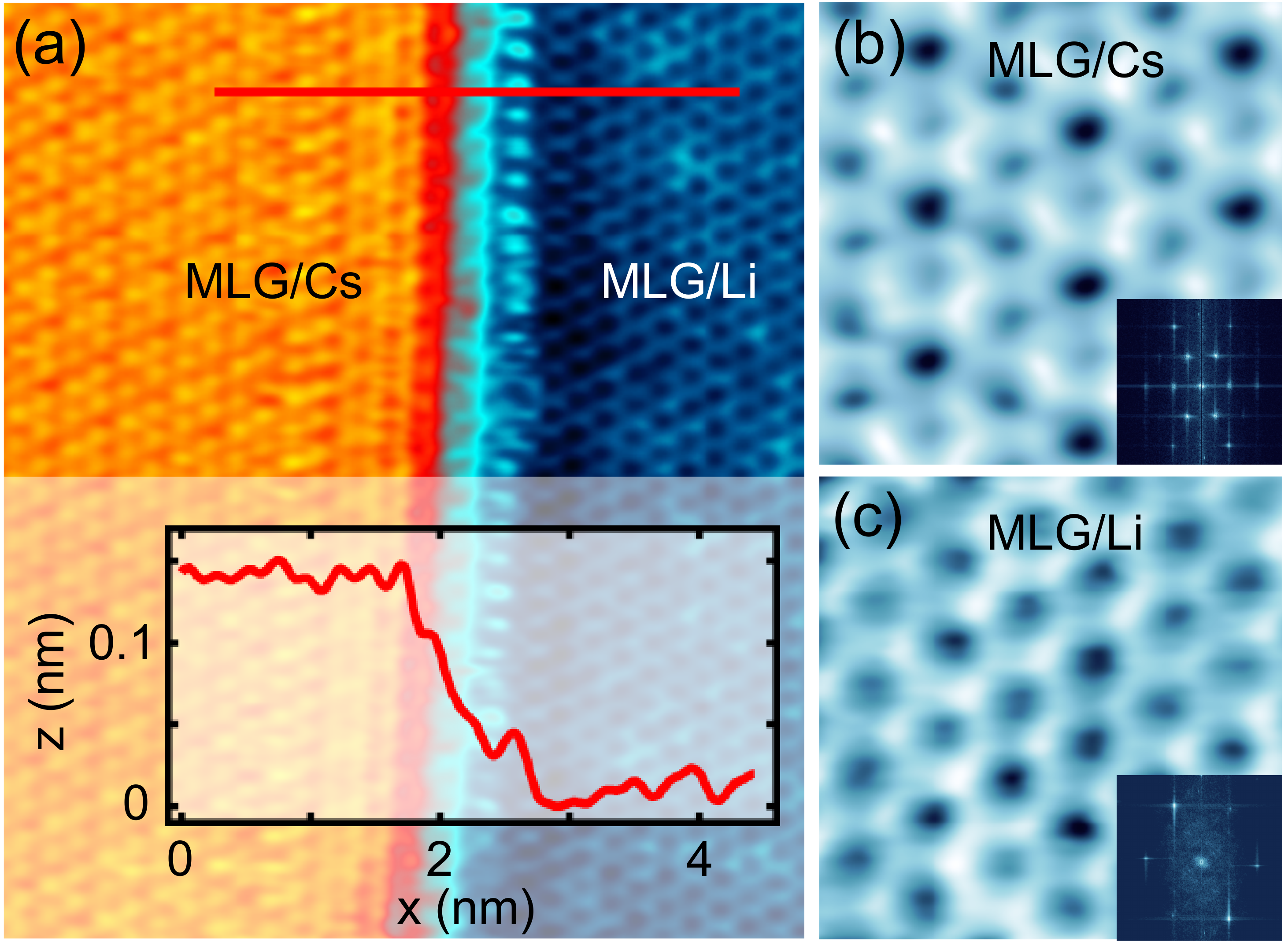}
\end{center}
\caption{(Color online) 
Atomic structure of Cs and Li intercalation phases under MLG on Ru(0001)\@.
(a) STM image of adjacent Cs and Li intercalation phases ($200\,\text{mV}$, $90\,\text{pA}$, $5.5\times 8\,\text{nm}^2$)\@.
Inset: Cross-sectional profile acquired along the line depicted in the top part of the STM image in (a)\@.
(b) Close-up view of Cs-intercalated domain with atomic resolution ($200\,\text{mV}$, $90\,\text{pA}$, $1.5\times 1.5\,\text{nm}^2$)\@.
Inset: Fourier transform of (b) ($11.9\times 11.9\,\text{nm}^{-2}$) with outer spots reflecting the C lattice of MLG.
The inner spots indicate the $(2\times 2)$Cs ordered array.
(c) Like (b) for a Li-intercalated region.
Inset: Like the inset to (b) with identical spots for the MLG C lattice and the $(1\times 1)$Li ordered structure.
}
\label{fig3}
\end{figure}

The alkali metal intercalation structures may be deduced from atomically resolved STM images.
Figure \ref{fig3}a presents the transition between MLG/Cs and MLG/Li regions with a difference in the apparent height of $\approx 0.13\,\text{nm}$ (inset to Figure \ref{fig3}a)\@.
The intercalant assemblies are best seen from the atomically resolved close-up views of Figure \ref{fig3}a.
In Figure \ref{fig3}b the honeycomb lattice of MLG is visible together with a regular hexagonal array of depressions.
As corroborated by the Fourier transform (inset to Figure \ref{fig3}b) the array reflects a $(2\times 2)$ lattice with respect to MLG\@.
Alkali metals were previously shown to preferrably adopt the central part of a C hexagon of graphite or graphene, \cite{prb_32_2538,nl_10_2828} which is assumed here for MLG/Cs as well.
This assumption is also reasonable with respect to the same conclusions drawn for a $(2\times 2)$Cs lattice observed from Cs-intercalated MLG on Ir(111)\@. \cite{natcommun_4_2772}
Due to the lattice mismatch between MLG and Ru(0001), the intercalated Cs atoms do not always occupy Ru on-top sites, which were demonstrated to be favored sites for the $(2\times 2)$Cs superstructure on clean Ru(0001)\@. \cite{prb_45_8638}

Intercalation of Li under MLG on Ru(0001) leads to STM images as presented in Figure \ref{fig3}c.
The MLG honeycomb lattice is visible only and a $(1\times 1)$Li assembly with respect to the MLG lattice is inferred (inset to Figure \ref{fig3}c); that is, each C hexagon is occupied by a Li atom.
An alternative explanation would be a disordered phase, which has recently been reported from room temperature photoemission experiments on Li-covered Ir(111) and Li-intercalated MLG on Ir(111)\@. \cite{prb_92_245415}
However, at $5\,\text{K}$, which is the temperature of the experiments reported here, the suggested ordered $(1\times 1)$Li intercalation phase is more plausible and agrees with the Li assembly observed at $6\,\text{K}$ for Li-intercalated MLG on Ir(111)\@. \cite{natcommun_4_2772}
For adsorption of Li on clean Ru(0001) a commensurate $(\sqrt{3}\times\sqrt{3})$R$30^\circ$-Li phase was discovered at low coverage with Li atoms preferrably residing at threefold hcp hollow sites of Ru(0001)\@. \cite{ss_337_198}
At larger submonolayer coverage incommensurate superstructures occur where Li atoms adopt different adsorption sites. \cite{prb_52_2927}

The intercalated $(2\times 2)$Cs and $(1\times 1)$Li phases observed here are formed with respect to the MLG lattice.
Therefore, the preferred adsorption sites on clean Ru(0001) are no longer energetically favored, which demonstrates the strong impact of graphene on the energy landscape for alkali metal adsorption. 

\section*{Conclusions}

Graphene can tune the atomic structure of intercalated phases.
The observed equilibrium assembly reflects the balance between couplings of the intercalant with graphene, the substrate and other intercalants.
Graphene on Pt(111) weakens the adsorption site specificity for intercalated Pt atoms and modifies the adsorption energy landscape.
The resulting surface reconstruction exhibits qualitative changes in the underlying dislocation network as well as an increased regularity and extension compared to reconstructions of pristine Pt(111)\@.
The intercalation of Cs and Li on graphene-covered Ru(0001) unravels a stronger impact of graphene on the intercalant assembly.
Alkali metal lattices form superstructures that are commensurate with the graphene lattice, irrespective of their substrate adsorption site.
The findings of this work may spark the tailoring of surface structures by the presence of graphene.

\section*{Acknowledgements}
Financial support by the Deutsche Forschungsgemeinschaft through Grants No.\,KR $2912/10-1$ and KR $2912/12-1$ is acknowledged. 

The authors declare no competing financial interest.


\providecommand{\latin}[1]{#1}
\makeatletter
\providecommand{\doi}
  {\begingroup\let\do\@makeother\dospecials
  \catcode`\{=1 \catcode`\}=2 \doi@aux}
\providecommand{\doi@aux}[1]{\endgroup\texttt{#1}}
\makeatother
\providecommand*\mcitethebibliography{\thebibliography}
\csname @ifundefined\endcsname{endmcitethebibliography}
  {\let\endmcitethebibliography\endthebibliography}{}

\end{document}